# Controlling Structure and Properties of Vapor-Deposited Glasses of Organic Semiconductors: Recent Advances and Challenges

*Kushal Bagchi [a], MD Ediger* [a]*

[a-]Department of Chemistry, University of Wisconsin-Madison, Madison, Wisconsin 53706, United States

AUTHOR INFORMATION

Corresponding Author: M.D Ediger

*Correspondence to: M.D Ediger

Phone: 608-262-7273

Email: ediger@chem.wisc.edu

## ABSTRACT:

The last decade has seen great progress in manipulating the structure of vapor-deposited glasses of organic semiconductors. By varying the substrate temperature during deposition, glasses with a wide range of density and molecular orientation can be prepared from a given molecule. We review recent studies that show the structure of vapor-deposited glasses can be tuned to significantly improve the external quantum efficiency and lifetime of OLEDs (organic light emitting diodes). We highlight the ability of molecular simulations to reproduce experimentally observed structures, setting the stage for in-silico design of vapor-deposited glasses in the coming decade. Finally, we identify research opportunities for improving the properties of organic semiconductors by controlling the structure of vapor-deposited glasses.

## TOC GRAPHICS

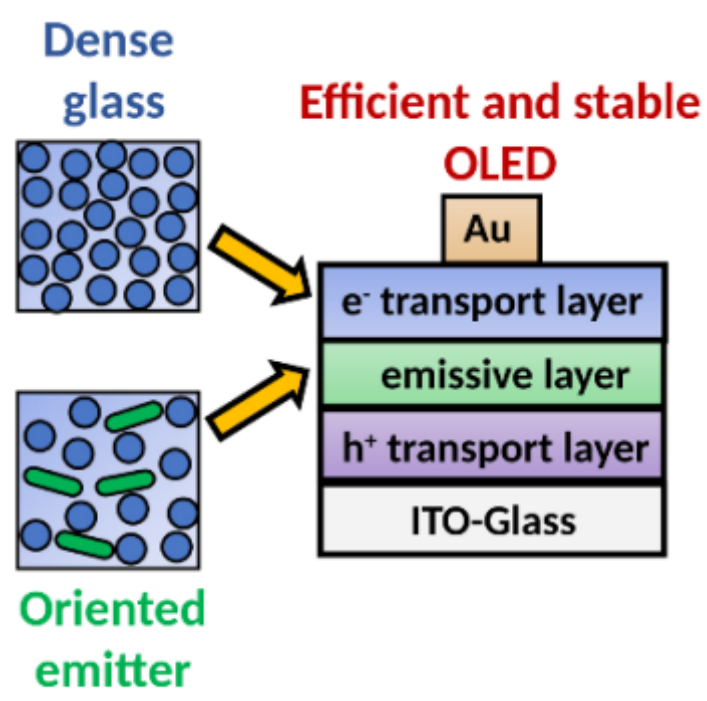

Organic semiconductors are an important class of materials for current and future technology. Organic semiconductors have already entered the electronic display market on a large scale. In 2018 more than half a billion OLED (organic light emitting diode) displays utilizing organic semiconductors were manufactured. Several commercially available consumer electronic items such as cellphones and televisions utilize these OLED displays. Organic semiconductors have also recently been commercialized for photovoltaics. In academic research labs organic semiconductors are being investigated for applications in next generation OLEDs[1] and photovoltaics[2], transistors[3] ,thermoelectric devices[4] and lasers[5]. In all these applications a major factor affecting the performance of an organic semiconductor is its physical state.

For certain applications, glassy solids offer key advantages over crystals. Glasses are compositionally more flexible than crystals; it is much easier to mix a dopant in the glassy state than in a crystal. In an OLED, the light emitting layer consists of a dilute emitter dispersed in a matrix; the *compositional flexibility* of the glassy matrix facilitates the preparation of the emissive layer. Molecular glasses tend to have smoother surfaces[6] compared to their polycrystalline counterparts, which is desirable for devices. Finally, unlike polycrystalline solids, glasses are free from grain boundaries, making them *macroscopically homogenous*. The macroscopic homogeneity of glasses is advantageous when uniformity is required over a large area. However charge conduction is less efficient in the glassy state as compared to the crystalline[7]; consequently glassy layers are particularly well-suited for applications, such as OLEDs which do not require high charge carrier mobility.

In organic electronic devices using *non-polymeric* organic semiconductors, glassy films are usually prepared either by solution processing or physical vapor deposition (PVD). A common solution processing method used in laboratory scale experiments is “spin-coating”; rapid spinning of a solution of the organic semiconductor causes evaporation of the volatile solvent and the formation of a glassy thin film. In PVD, the material of interest is evaporated or sublimed in a vacuum chamber. The evaporated (or sublimed) material condenses onto a substrate to form a glassy thin film. PVD has long been the standard route to prepare glassy layers in OLEDs[8]. PVD is also used to fabricate amorphous films for organic photovoltaic (OPV) cells[9,10,11] and in a few cases has also been used for OFET[12] (organic field effect transistors) fabrication. However many of the advantages PVD offers over solution processing have been appreciated only more recently[13]. Moreover, an important influence of the temperature at which the layers of an OLED are deposited on device performance has been demonstrated quite recently[14].

In this perspective, we explore the relationship between glass structure and material properties for vapor-deposited glasses, in the context of organic electronic devices. Although we consider other applications, we focus our attention on OLEDs, as they are the most prominent technology utilizing vapor-deposited glasses. PVD glasses are well-suited for studying structure-property relations for devices because a broad range of structures can be prepared from a given molecule by changing the deposition temperature[15] and rate[13,16].

In this article, we first review general principles regarding the anisotropic structure of PVD glasses. We then discuss emitter orientation and dipolar order, aspects of anisotropic glass structure directly related to OLED device performance. We next examine how the tight and dense packing of PVD glasses can be used to enhance charge carrier mobility and performance of OLEDs. Finally, we describe new research opportunities in the field. Throughout this perspective, we discuss the promise of computational design as a means to optimize PVD glasses of organic semiconductors.

**Anisotropic structure of vapor-deposited glasses:** In the following seven paragraphs, we review general features of the anisotropic structure of PVD glasses, providing background for subsequent portions of the article.

*A broad range of anisotropic glassy structures are accessible by physical vapor-deposition*. Shown in **Figure 1** are two-dimensional X-ray scattering patterns from one spin-coated glass and three vapor-deposited glasses. A corresponding schematic of the structure in the film is shown below each diffraction pattern. Beneath the schematic, the structure of the molecule deposited is depicted. The diffraction patterns were collected in grazing incidence, which is suitable for studying the structure of thin films[17]. The out-of-plane structure of the glass determines the scattering along $Q_z$ while the in-plane structure determines the scattering along $Q_{xy}$. In contrast to the spin-coated film, all the vapor-deposited glasses in **Figure 1** exhibit anisotropic scattering features. That is, for the PVD glasses, the scattered intensity is different along different directions. For the spin-coated glass, the scattered intensity is roughly the same along all directions; this indicates that the structure is isotropic, as traditionally expected for glasses. Consistent with **Figure 1**, dichroism studies by Yokoyama and co-workers have found that PVD glasses are structurally anisotropic, while spin-coated films of the same molecules typically exhibit isotropic structure[13]. All the diffraction patterns shown in **Figure 1** were obtained from glasses prepared by depositing onto a substrate held at a temperature below the glass transition temperature. Depositing above $T_g$ usually yields isotropic glasses, with similar properties as liquid-cooled glasses.

*Vapor-deposited glasses can exhibit translational order*. The two left-most X-ray scattering patterns in **Figure 1** are from glasses of $Alq_3$ (Tris(8-hydroxyquinolinato) aluminum) prepared by different routes; one of the films was spin-coated and the other vapor-deposited. $Alq_3$ is a common electron transport and light emitting layer in OLED devices ($Alq_3$ was used in the first thin-film

OLED ever fabricated[18]). The PVD glass of $Alq_3$, deposited at 280 K ($0.62T_g$), exhibits a broad anisotropic scattering feature in the out-of-plane direction at $Q_z \approx 0.8$ Å$^{-1}$; in real space this corresponds to a distance of ≈ 8 Å, which is roughly the molecular diameter of $Alq_3$. Anisotropic scattering at the length-scale of the molecular diameter can be interpreted as a tendency towards molecular layering, as indicated in the accompanying schematic. Simulations have shown that this type of scattering feature arises from center of mass correlations along the surface normal[19,20]. Molecular layering is a common feature in PVD glasses[21,22,23].

*Vapor-deposited glasses can also exhibit orientational anisotropy*. The two right-most patterns in **Figure 1** are obtained from vapor-deposited glasses of posaconazole and DSA-Ph (1-4-Di-[4-(N,N -diphenyl)amino]styryl-benzene) deposited at $0.99T_g$ and $0.81T_g$ respectively. DSA-Ph is used as a blue-light emitter in OLED devices. Posaconazole is not an organic semiconductor, but is an important model system which provides useful insights into the process of vapor-deposition[21,24]. Both these glasses exhibit anisotropic scattering at $Q \approx 1.4$ Å$^{-1}$. In the vapor-deposited glass of posaconazole there is higher scattered intensity in the plane (along $Q_{xy}$) and for DSA-Ph there is greater intensity out of the plane (along $Q_z$). The posaconazole and DSA-Ph glasses exhibit strong tendencies towards vertical and horizontal molecular orientation respectively as depicted in the schematics. The strong tendency towards horizontal molecular orientation observed for DSA-Ph has been observed for several different organic semiconductors deposited at $0.7\text{-}0.8T_g$[25]. There is evidence that horizontal molecular orientation is desirable for OLED devices[26,8].

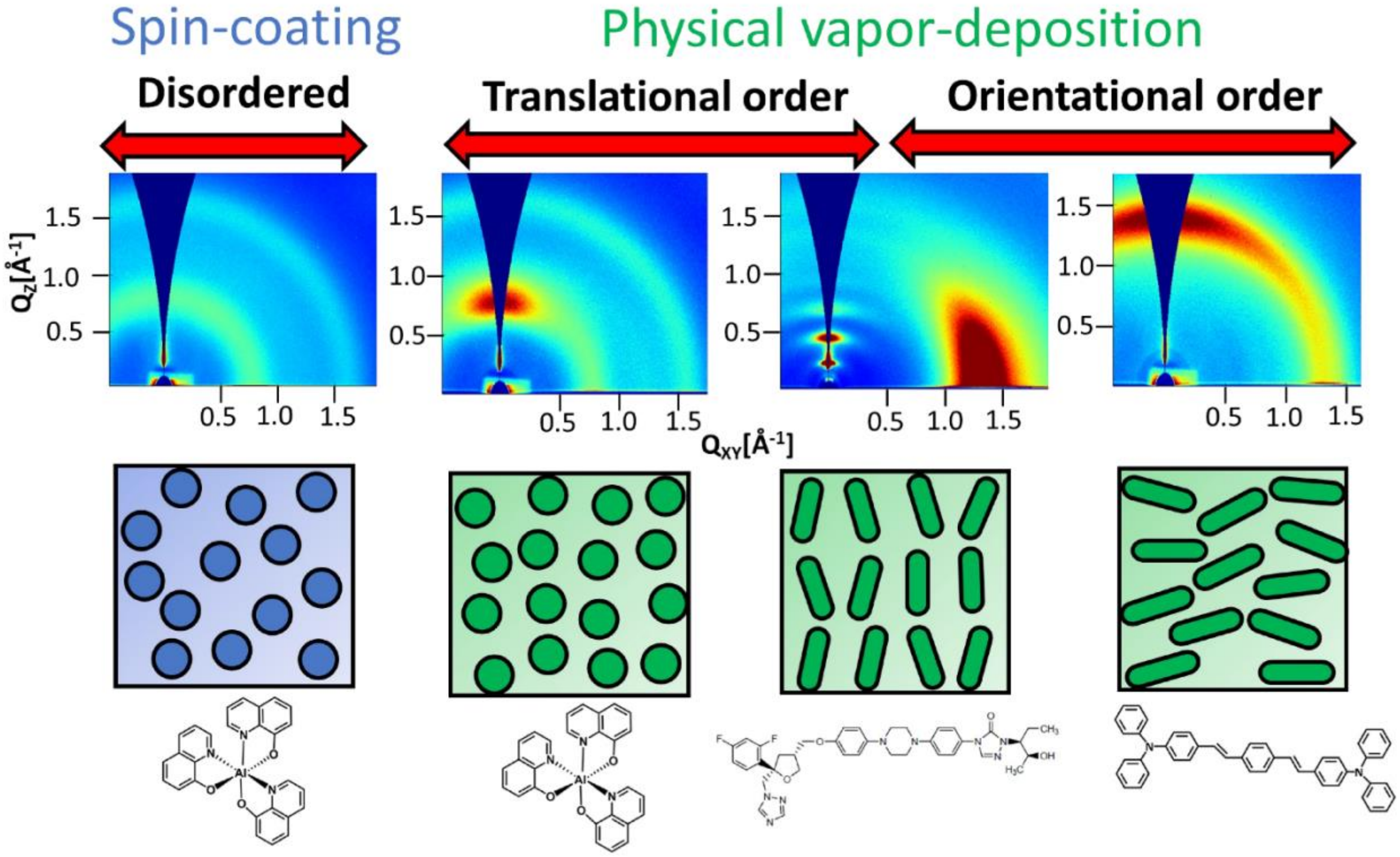

**Figure 1**: X-ray scattering patterns (from left to right) of spin-coated $Alq_3$ and vapor-deposited $Alq_3$, posaconazole, and DSA-Ph glasses. $Alq_3$ was deposited at 280 K($0.62T_g$), posaconazole at 328 K($0.99T_g$), and DSA-Ph at 290 K($0.81T_g$). The colors in the diffraction patterns represent scattered intensity; for a given pattern, red represents high scattered intensity and blue represents low-scattered intensity. The spin-coated glass does not exhibit any anisotropic scattering features, whereas the vapor-deposited glasses exhibit anisotropic scattering features that arise from translational and/or orientational order. Simplified illustrations depict the structure associated with each X-ray scattering pattern. The molecular structure of each of the molecules is shown below the schematics. Measurements were performed by Ediger and co-workers. The $Alq_3$ patterns are reprinted with permission from reference 20. Copyright 2018 American Chemical Society. The posaconazole pattern is reproduced with permission from reference 21. The DSA-Ph pattern is reproduced with permission from reference 25. Copyright 2019, Royal Society of Chemistry (Great Britain).

*Vapor-deposition can be used to prepare glasses with liquid crystalline order.* The GIWAXS pattern in **Figure 1** shows that posaconazole, prepared at 328 K, exhibits both the high degree of translational and orientational order characteristic of aligned smectic liquid crystals[21]. Remarkably, posaconazole does not have equilibrium liquid crystal phases. This type of highly organized smectic packing can be advantageous in OFETs[27,28].

*A wide range of different glassy structures can be obtained from the same molecule by selecting the substrate temperature used for deposition*. Shown in **Figure 2A** is the orientational order parameter, $S_z$, as a function of the deposition temperature for vapor-deposited glasses of the common hole transport material TPD (N,N′-bis(3-methylphenyl)-N,N′-diphenylbenzidine). The X-axis, $T_{substrate}/T_g$, is the deposition temperature normalized by the glass transition temperature of TPD. The order parameter is determined from variable angle spectroscopic ellipsometry (VASE). The orientation order parameter, $S_z$, quantifies the molecular orientation of the long axis of TPD. Orientation order parameters of -0.5, 1.0 and 0 correspond to perfectly horizontal, perfectly vertical and completely random orientation of the molecular long axis respectively. Depositing TPD between 0.7-0.8$T_g$ produces glasses with a strong tendency for horizontal molecular orientation, while depositions at ≈ 0.95$T_g$, produce films with a weak tendency for vertical molecular orientation. **Figure 2** shows that vapor deposition enables structure to be controlled extremely precisely. By picking the appropriate substrate temperature, a macroscopic glassy structure with any specific average molecular orientation from -0.4 to +0.1 can be chosen. The average molecular orientation influences important properties of vapor-deposited glasses. For PVD glasses of TPD, the elastic modulus[29] and in-plane thermal conductivity[30] correlate with molecular orientation.

Computer simulations successfully predict the experimentally observed trend in molecular orientation with deposition temperature. Shown in **Figure 2B** is the orientational order parameter, $S_z$, as a function of deposition temperature, obtained from computer simulations of the deposition process for a coarse-grained model of TPD. Similar to what is seen experimentally, depositions at lower substrate temperature produce glasses with horizontally oriented molecules, while depositions closer to $T_g$ produce films with vertical molecular orientation. The same trend has been observed in simulations of vapor deposition that utilized an atomistic model of TPD[31]. As we discuss below, the ability of simulations to predict experimental trends opens the exciting possibility for in-silico design of vapor-deposited glasses in the coming decade.

Molecular simulations indicate that a "surface equilibration" mechanism controls structure formation in vapor-deposited glasses. While we direct the interested reader to a recent review[32] that explains this mechanism in detail, we summarize here the central ideas. Surface diffusion measurements have shown that mobility at the surface of an organic glass can be up to 8 orders of magnitude higher than mobility in the bulk[33]. Thus, during vapor-deposition onto a substrate held below $T_g$, the surface can be fluid while the bulk is immobile. Computer simulations have shown that molecules at the free surface are mobile enough to equilibrate towards configurations favored at the surface of the supercooled liquid[15]. Upon being buried by subsequent deposition these molecules lose their mobility, resulting in surface-favored configurations being trapped into the bulk of the glass. The surface equilibration mechanism therefore predicts that under conditions where there is enough surface mobility, the bulk structure of a vapor-deposited glass will resemble the surface structure of the supercooled liquid. The surface equilibration mechanism correctly predicts the structure of vapor-deposited glasses of $Alq_3$[20], posaconazole[21], and TPD[15].

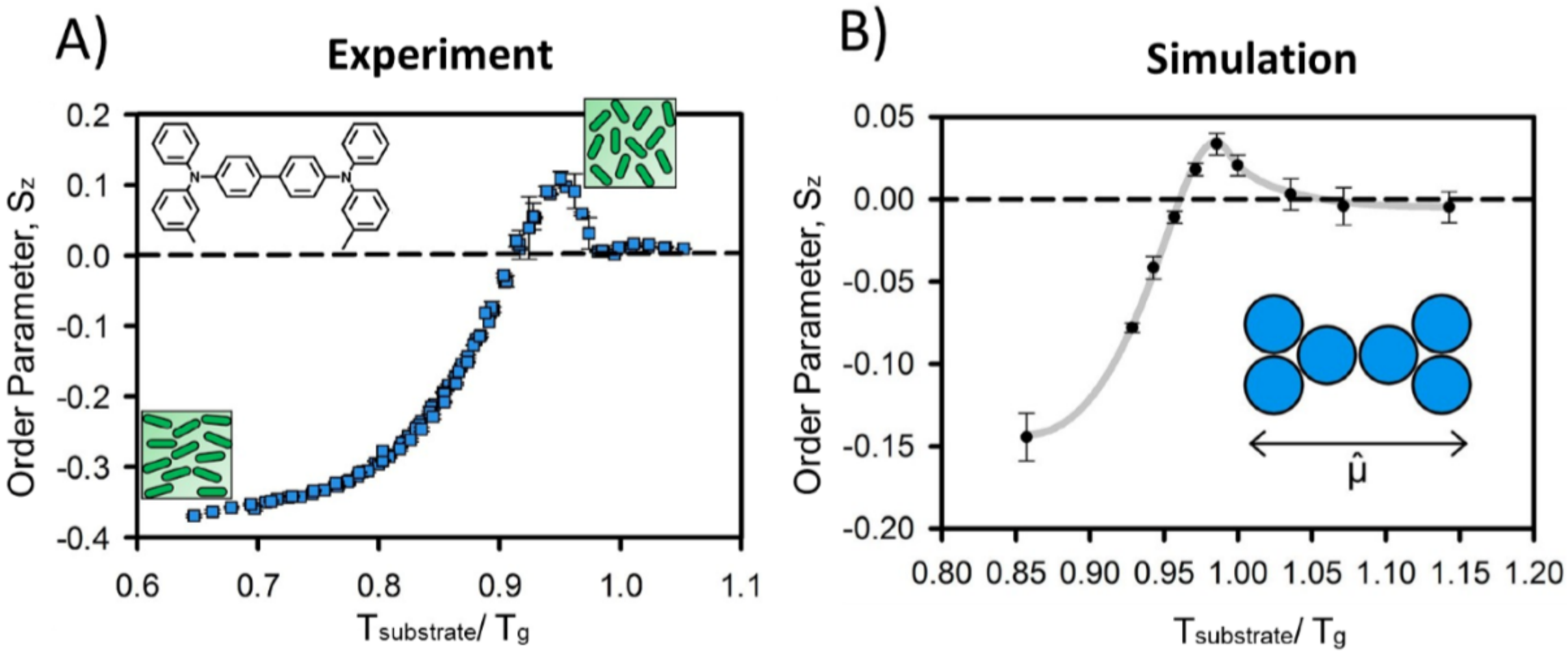

**Figure 2**: The orientational order parameter, $S_z$, for the long axis of vapor-deposited glasses of TPD (shown in inset) observed in experiments(A) and in simulations of a coarse-grained representation of TPD(B). The experiments and simulations show that molecular orientation can be tuned by selecting the substrate temperature used for deposition. Substrate temperatures of 0.65-$0.8T_g$ produce glasses with horizontal molecular orientation while glasses prepared at $\approx 0.95T_g$ exhibit vertical molecular orientation. Molecular orientation is measured at room temperature for all the glasses. Simplified schematics of lozenges depict horizontal and vertical molecular orientation in glasses deposited at low and high substrate temperature, respectively. Simulations of the vapor deposition process using a coarse-grained Lennard-Jones-representation of TPD (shown in the inset of B) reproduces the experimental trends in the order parameter. Each bead of the coarse-grained model represents an aromatic ring in TPD. Experiments were performed by Ediger and co-workers and simulations were performed by de Pablo and co-workers. Reproduced with permission from reference 15.

**Emitter orientation:** The emissive layer in an OLED device consists of a dilute (1- 10 wt %) emitter dispersed in a host matrix. The molecular orientation of the emitter determines how much of the emitted light escapes from the device. A horizontal orientation of the emissive transition dipole moment vector of the emitter molecule results in greater light out-coupling from an OLED device compared to vertical and isotropic orientations. This is because a horizontal orientation of the emissive transition dipole moment vector minimizes power dissipation to waveguide and surface-plasmon modes. Maximizing horizontal orientation of the emissive transition dipole moment vector is an area of extensive research within the OLED community. Brütting and co-workers have identified emitter orientation as a key variable influencing the efficiency of an OLED[26].

The orientation of a molecule in a vapor-deposited mixture, like in a single component PVD glass, can be controlled by substrate temperature during deposition. Shown in **Figure 3** is the orientational order parameter $S_z$, of the emitter Coumarin 6 co-deposited with various hosts, plotted as a function of $T_{substrate}/T_{g,host}$[34]. The orientational order parameter of the emitter is determined from angle-dependent photoluminescence measurements. Rather than changing the substrate temperature during deposition the authors[34] co-deposit coumarin 6 with host molecules that have different glass transition temperatures. Although all depositions are performed at room temperature, by changing the host, the authors change the ratio of $T_{substrate}/T_g$[34]. The trend seen in the orientation of a dilute emitter seen in **Figure 3** is strikingly similar to the trend seen for molecular orientation in a single component vapor-deposited glass in **Figure 2.** This strongly suggests that the same factors can influence molecular orientation in both pure and two-component PVD glasses.

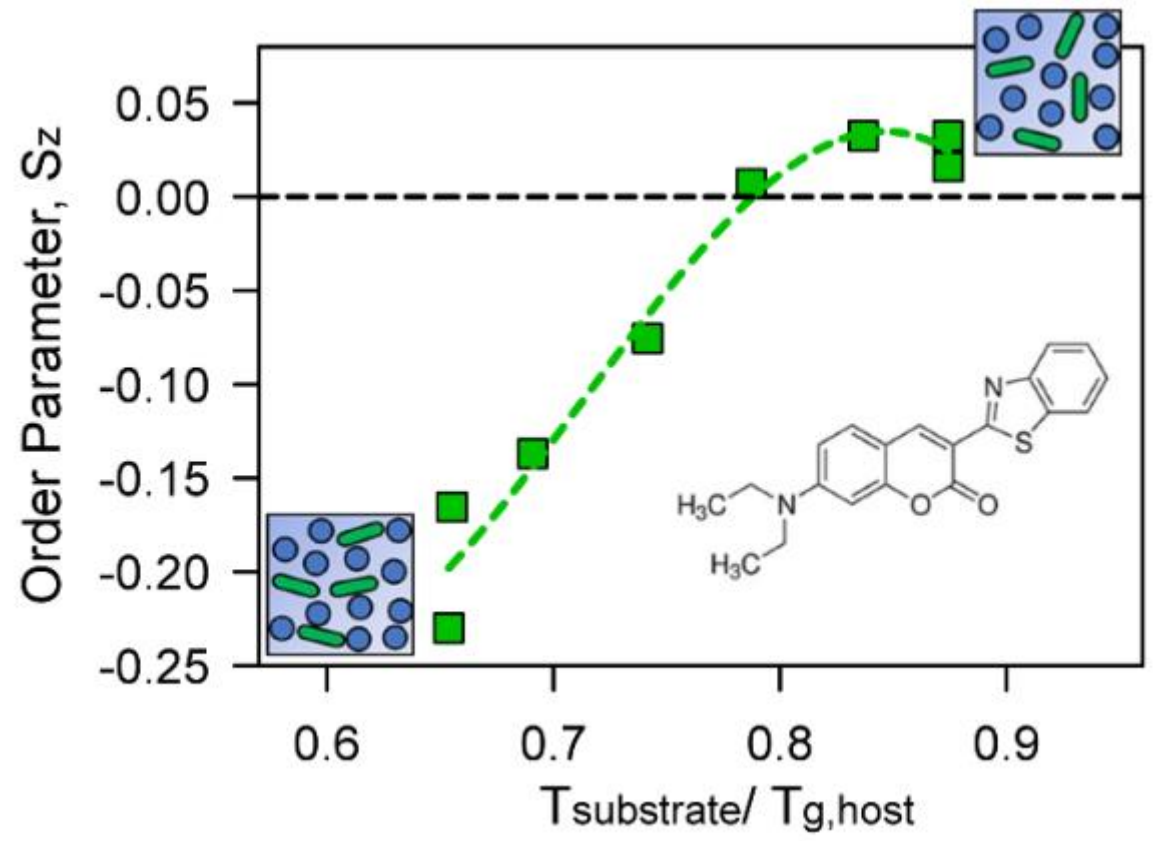

**Figure 3:** The orientational order parameter, $S_z$, of a dilute emitter, Coumarin 6 (shown in inset) co-deposited in matrices with different glass transition temperatures $T_g$. The concentration of the emitter is 2% vol. The light emitting layer in OLED devices is a usually a dilute emitter in a glassy matrix. For efficient light outcoupling from an OLED device, horizontal molecular orientation of the emissive transition dipole moment is desired. Horizontal molecular orientation is achieved at low $T_{sub}/T_g$. For this co-deposited system, molecular orientation is controlled by the normalized deposition temperature, similar to results for single component systems (Figure 2). Measurements were performed by Brütting and co-workers. Adapted with permission from reference 34. Copyright 2015 American Chemical Society.

Qualitatively similar trends in emitter orientation to that seen in **Figure 3** have also been observed for phosphorescent[14] and TADF (thermally activated delayed fluorescence) emitters[35]. Adachi and coworkers observed horizontal orientation ($S_z = -0.31$) of the emitter at a deposition temperature of 200 K and nearly isotropic orientation ($S_z=0.05$) when the deposition was performed near room temperature (300 K). The authors found a correlation between the orientation of the emitter and OLED device performance. The device in which the emitter molecule was horizontally aligned had a 24% higher external quantum efficiency (EQE- the number of photons emitted per injected charge carriers) compared to the device with nearly isotropic emitter orientation. In this study, the different alignments of the emitter were attained by changing the deposition temperature of the emissive layer.

While the importance of the deposition temperature (relative to the glass transition temperature) in influencing emitter orientation has been clearly demonstrated, the role of specific host-guest interactions is less clear. Based on molecular simulations, Moon et al[36] attribute the alignment of phosphorescent iridium emitters to van der Waals and electrostatic interactions with the host. For a different class of emitters (TADF emitters), based on measurements of emitter orientation, and electrical measurements, Adachi and co-workers[37] discuss the role of host-guest dipole-dipole interactions on emitter orientation. In these studies, systems are compared at different $T_{substrate}/T_{g,host}$ in some cases, while in other cases different emitters are deposited in the same host. Such comparisons make it difficult to disentangle the role of host-guest interactions on emitter alignment from the influence of the deposition temperature or the inherent propensity of an emitter to exhibit anisotropic orientation. Experimental comparisons utilizing the same emitter, deposited in different hosts but at the same $T_{substrate}/T_{g,host}$, would best elucidate the role of host-guest interactions in determining emitter orientation.

We find two examples in the literature which can help us isolate the role of host-guest interactions on emitter orientation; these studies allow comparison of emitter orientation in different hosts at same $T_{substrate}/T_g$. The first, a study by Jing et al[38] suggests that host-guest interactions are less important compared to $T_{substrate}/T_{g,mixture}$ in determining molecular orientation. The orientational order parameter of DSA-Ph, in a single component and in mixtures with 20 to 80% $Alq_3$, collapse fairly well onto the same curve when plotted as a function of $T_{substrate}/T_{g,mixture}$. The average orientation of DSA-Ph molecules, whether they are surrounded by $Alq_3$ or other DSA-Ph molecules is determined primarily by deposition temperature relative to the glass transition of the mixture. At higher substrate temperatures ($T_{substrate}/T_{g,mixture} \approx 0.95$), there is a slight composition dependence of the orientational order parameter of DSA-Ph, suggesting that additional factors such as host-guest interactions may also play a role. The second example of such a study can be seen in **Figure 3,** at $T_{substrate}/T_{g,host} \approx 0.65$. When compared at the same $T_{substrate}/T_g$, the tendency for horizontal orientation of the emitter Coumarin 6 (shown in **Figure 3**) is ≈ 1.4 times greater in an $Alq_3$ matrix than a spiro-CBP matrix. It is plausible that host-guest interactions are responsible for this difference in emitter orientation in the two hosts. While Coumarin 6 and $Alq_3$ are highly polar molecules with dipole moments of ≈ 5.8 D and ≈ 4 D respectively, Spiro-CBP has almost no dipole moment[34].

The dependence of emitter orientation on substrate temperature is now reasonably well understood. On the other hand, the dependence of emitter orientation on deposition rate has not been explored, to the best of our knowledge. Several studies of single-component PVD glasses have shown that anisotropic structure can also be manipulated by changing deposition rate[13,21,39,40]; at a given temperature, a lower deposition rate allows further equilibration towards the structure preferred at the surface. By analogy, deposition rate is expected to affect emitter orientation and such a relationship might be harnessed for improving device performance.

**Surface potential:** Vapor-deposition can produce glasses with dipolar order. That is, the *dipole moment vector* of a molecule can, on average, be preferentially oriented in a PVD glass. When the dipole moment vector has a preferred orientation, the interfaces of the vapor-deposited glass become charged. In OLED devices interfacial charges can modulate charge injection barriers[41] and cause exciton quenching[42,43], consequently influencing the performance of the device. In vibration-based electret generators, high levels of dipolar order are advantageous[44]. The dipole moment vector (determined by the ground state charge distribution) should not be confused with a transition dipole moment vector discussed above (which is the vector associated with a given spectroscopic transition); these two vectors can point in different directions in a molecule and can consequently have different net orientations in a film.

Shown in **Figure 4** is the surface potential of several common organic semiconductors as a function of film thickness; measurements are performed with a Kelvin probe. The alignment of the dipole moment vector results in the production of a surface potential that grows linearly as a function of film thickness. The measured surface potential depends on the extent of alignment of the average dipole moment vector in the *bulk* of the film. The surface potential, despite what the name suggests, is a measure of bulk glass structure. The surface potential also depends on the magnitude of the dipole moment vector. While the surface potential of polar organic semiconductors ($Alq_3$, TPBi,OXD-7 and BCP) in **Figure 4** increases significantly as a function of film thickness, for non-polar organic semiconductors (UGH-2 and CBP) the surface potential appears flat on the same plot. A positive surface potential implies a positively charged free surface and negatively charged buried interface. It is important to point out that only a very small degree of alignment is required to produce a large measured surface potential. For $Alq_3$, a 2 % alignment (experimentally measured polarization is 2% of the calculated polarization for perfect perpendicular orientation of dipole moment vector relative to surface) of the dipole moment vector gives rise to a giant surface potential of ≈ 5 V at 100 nm[45]. Similar to other aspects of PVD glass structure, dipolar ordering can also be predicted from computer simulations[46].

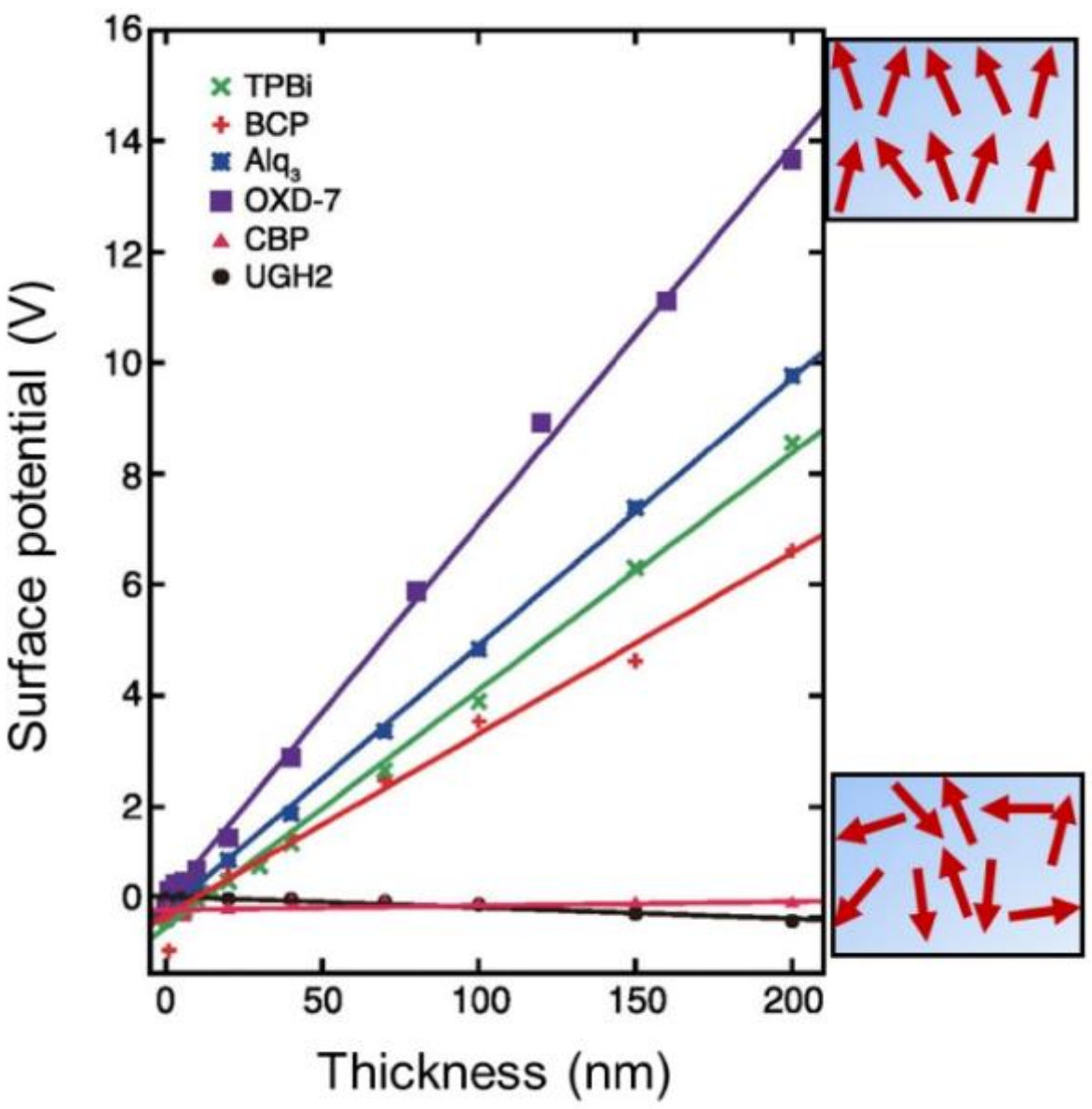

**Figure 4**: The surface potential of vapor-deposited glasses of organic semiconductors measured by Kelvin probe measurements. As a result of dipolar order, the surface potential of glasses of polar molecules increases linearly as a function of film thickness. Measurements were performed by Ishii and co-workers. Adapted with permission from reference 47. Copyright 2012, AIP Publishing.

Almost all organic semiconductors that exhibit a large surface potential when vapor-deposited as glasses, exhibit a *positive* surface potential. The general tendency for vapor-deposited organic semiconductors to produce positive surface potentials remains to be understood. An interesting exception to this general rule is the vapor-deposited glass of $Al(7\text{-}Prq)_3$, an alkylated derivative of $Alq_3$, which exhibits a negative surface potential[41]. The surface equilibration mechanism[32] (described briefly above), which makes predictions about the bulk structure of a vapor-deposited glass based on the surface structure of the supercooled liquid, correctly predicts the positive surface potential of $Alq_3$. This comparison requires input from molecular dynamics computer simulations, and it would be useful to check if this approach also correctly predicts the negative surface potential of $Al(7\text{-}Prq)_3$.

To understand the influence of dipolar orientation on performance, Ishii and co-workers fabricated OLED devices using $Alq_3$ and $Al(7\text{-}Prq)_3$ as electron transport layers (and NPD as the hole transport layer)[48]. Both layers were deposited at room temperature. The top surfaces of $Alq_3$ and $Al(7\text{-}Prq)_3$ are positively and negatively charged respectively; the cathode which injects electrons into the device is deposited on top of the electron transport layers. In the device with the $Al(7\text{-}Prq)_3$ as the electron-transport layer there was evidence of an increased electron injection barrier, which is undesirable in an OLED device. The higher electron injection barrier for $Al(7\text{-}Prq)_3$ is a result of its negatively charged top surface (owing to the orientation of the dipole moment vector). The device using $Alq_3$ as the electron transport layer exhibited a 65% higher maximum luminous efficacy (LE); the luminous efficacy is the luminous flux divided by power.

Brütting and co-workers have recently demonstrated that doping molecules that exhibit dipolar ordering into charge transporting layers that do not exhibit spontaneous polarization can be used to influence the charge injection characteristics of the latter. Under certain circumstances, this type of “dipolar doping” can lead to an order of magnitude enhancement in the current density in an OLED device[49].

Dipole-dipole interactions apparently play an important role in determining the dipolar order in vapor-deposited mixtures of organic semiconductors. In comparison to a neat film, dipole alignment of the polar molecule Alq3 was shown to be up to six times higher when co-deposited into a host matrix of the non-polar material NPD[50]. This result can be rationalized if dipole-dipole interactions in the neat Alq3 are so strong that they hinder overall alignment.

Several issues related to surface potentials in PVD glasses of organic semiconductors could benefit from further investigation. For example, the dependence of dipolar ordering of PVD glasses on deposition conditions is almost unexplored. A recent study by Holmes & co-workers[43] on PVD glasses of electron transport material TPBi (2,2′,2"-(1,3,5-Benzinetriyl)-tris(1-phenyl-1-H-benzimidazole)) concludes that dipolar ordering decays monotonically with deposition temperature, with the net orientation of the dipole moment vector becoming isotropic at ~$0.9T_g$. Since reducing the deposition rate generally has a similar effect on PVD glass structure as increasing the deposition temperature[13,16], it is expected that dipolar ordering can also be modulated by deposition rate.

**Charge mobility:** The structure of a molecular glass significantly influences its charge carrier mobility. Yokoyama et al showed that a PVD glass of BSB-Cz (4,4′-bis[(*N*-carbazole)styryl]biphenyl) deposited at 0.77 $T_g$ can exhibit approximately ≈ 5 times higher electron mobility than a glass deposited at 0.98 $T_g$. Like TPD (**Figure 2**), the BSB-Cz glass deposited at $0.77T_g$ exhibits much more pronounced horizontal molecular orientation compared to the glass prepared at 0.98 $T_g$. Yokoyama et al attribute the five-fold enhancement in mobility to the better electronic coupling associated with horizontal molecular orientation[51]. A similar argument was used to rationalize better charge conduction in an anisotropic vapor-deposited glass of TCTA (Tris(4-carbazoyl-9-ylphenyl)amine) compared to an isotropic spin-coated glass of the same molecule[52]. However, changing the substrate temperature during vapor-deposition or the preparation route (spin-coating vs vapor-deposition) changes the film density as well as the molecular orientation in the glass. In principle, both a higher density and horizontal molecular orientation can improve charge transport out of the plane. Until recently the role of film density on charge transport in glasses has been overlooked and enhancements in charge transport have been attributed to horizontal molecular orientation.

Recent studies from the Adachi group demonstrate that both film density and molecular orientation need to be considered when rationalizing differences in charge conduction between glasses with different preparation histories. Shown in **Figure 5** are the charge carrier mobility, film density and orientational order parameter of vapor-deposited glasses of α-NPD (*N*,*N*′-di(1-naphthyl)-*N*,*N*′-diphenyl-(1,1′-biphenyl)-4,4′-diamine) as a function of deposition temperature. The plot shows that the α-NPD glass deposited between 0.7-0.8$T_g$ can have a (zero-field) charge carrier mobility roughly ≈ 25 times higher than the glass deposited at 0.6$T_g$ [53]; the charge mobility enhancement is accompanied by a 1% density difference. The authors conclude that charge carrier mobility for PVD glasses of α-NPD correlates better with film density than the orientation order parameter.

Studies so far have focused on correlating either horizontal molecular orientation or enhanced density with charge transport in PVD glasses. However, there can be other aspects of glass structure that could also influence charge transport. Earlier in the perspective we briefly discussed molecular layering, which is a common packing motif in PVD glasses. The influence of molecular layering on charge transport is not known. Additionally, for polymeric semiconductors, the orientational correlation length, extracted from resonant soft X-ray scattering, has been shown to correlate with charge carrier mobility[54]. The orientational correlation length is defined as the shortest distance it takes for the director orientation to go from vertical to horizontal orientation (or vice-versa) in a material with liquid-crystalline order. The orientational correlation lengths of anisotropic vapor-deposited glasses have not been measured. Moreover, studies till now have focused on correlating the bulk structure of vapor-deposited glasses to charge transport. Interfacial structure influences charge injection barriers in OLEDs and charge conduction in OFETs. No study exists, to the best of our knowledge, that correlates the interfacial structure of a vapor-deposited glass to charge injection or conduction.

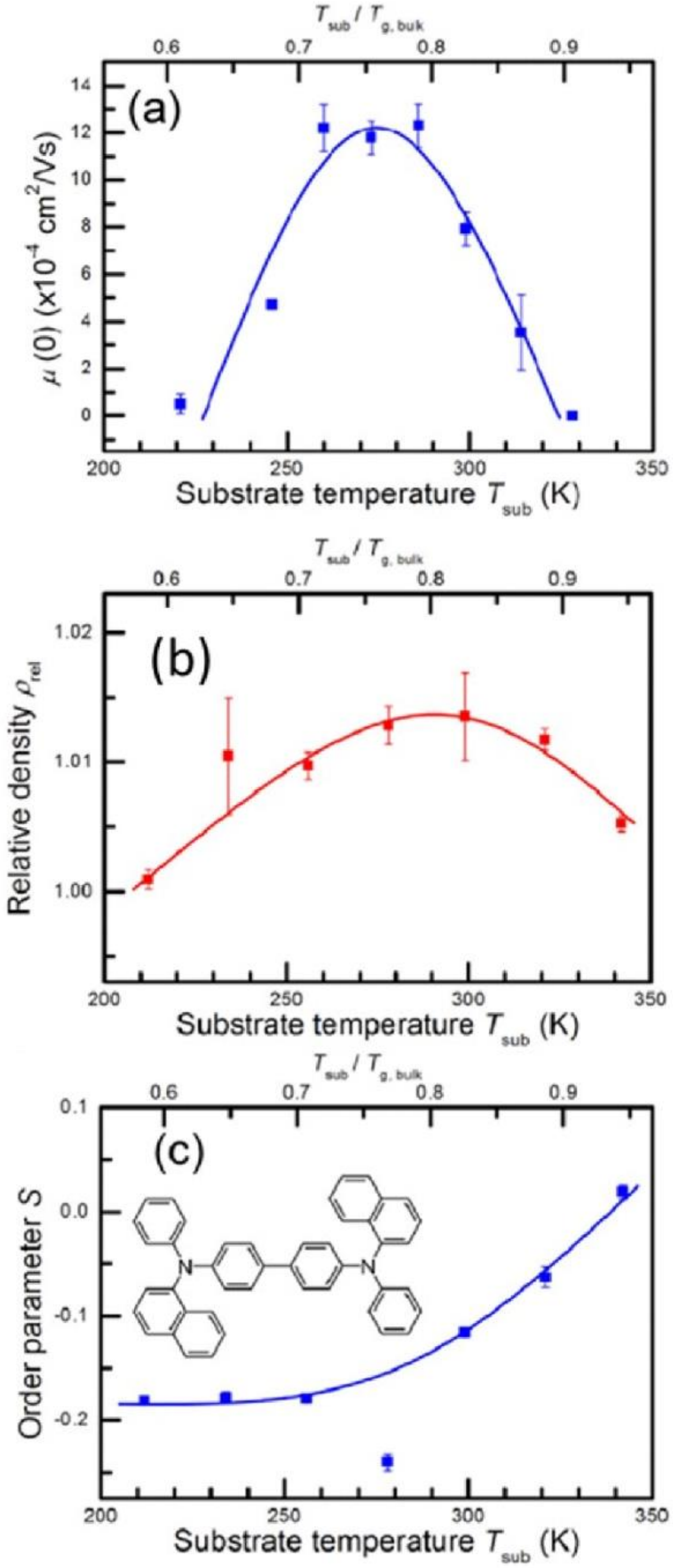

**Figure 5**: Zero-field charge carrier mobility (a), relative density (b), and orientational order parameter $S_z$ (c), plotted as a function of deposition temperature for vapor-deposited NPD glasses. The molecular structure of NPD is shown in the inset of (c). Charge carrier mobility seems to correlate better with film density than with molecular orientation. Measurements were performed by Adachi and co-workers. Reprinted with permission from reference 53. Copyright 2017 American Chemical Society.

**OLED device performance:** Improving the efficiency and lifetime of OLED devices is an extensive area of research today in the organic electronics community. Recent studies have demonstrated that the substrate temperature during deposition can be tuned to improve OLED performance[14,35,43]. While OLEDs are typically fabricated at room temperature, Reineke and co-workers[14], prepared devices with the electron-transport (ETL) and emissive layer (EML) deposited across a range of deposition temperatures. The authors found that depositing at $\approx 0.85T_g$ results in significant improvements in OLED performance. Shown in **Figure 6** is the external quantum efficiency (EQE) and luminous efficacy (LE) of a green light emitting OLED as a function of deposition temperature (of the ETL and EML layer). The LE and EQE for the device fabricated at $\approx 0.85T_g$ improved by 37% and 24% respectively compared to the device prepared near room temperature. For three other devices using the same host and charge transport layers but different emitters, deposition at the optimal temperature window resulted in EQE enhancements from 15% to 163 % relative to devices fabricated at room temperature. The authors show these enhancements in EQE are a result of the improved radiative efficiency of the emitter in the glasses prepared at $\approx 0.85T_g$[14]. Glasses prepared at this deposition temperature are expected to be maximally dense and the authors hypothesize that the highly efficient packing around the emitter limits radiationless relaxation pathways. In subsequent work**,** Holmes and co-workers, prepared an OLED device with the same charge transport layers as **Figure 6** but with a different emitter(Ir(ppy)3); they report that depositing the EML and ETL layers at $\approx 0.85T_g$ results in 18% enhancement in EQE relative to room temperature[43]. In contrast to the work of Reineke and co-workers, Holmes and co-workers attribute the enhancement in EQE to reduced dipolar order in glasses deposited at higher temperatures.

Changing the deposition temperature of an emissive or charge transport layer in an OLED changes many functionally important aspects of PVD glass structure simultaneously: film density[15], dipolar order[43]and orientation of emissive transition dipole moment vector[35]. These different aspects of glass structure may be optimized for performance at different deposition temperatures. For instance, **Figure 3** shows that emitter orientation is optimized at ~0.65$T_g$ while it is well known that density is optimized at ~0.85$T_g$. The relative importance of these different aspects of structure will vary based on the molecule or combination of molecules used to construct a given layer in an OLED. The ideal deposition temperature, relative to $T_g$, for a layer in an OLED is therefore likely to vary from system to system.

For the device shown in **Figure 6**, depositing in the optimal substrate temperature window also resulted in a five-fold improvement in device lifetime compared to deposition at room temperature. Adachi and co-workers also observe that optimizing density (by picking the appropriate deposition temperature) can improve air stability of α-NPD diodes over extended periods of time[53]. Enhancements in device lifetime have been discussed in the context of the greater chemical stability of the dense PVD glasses prepared at ≈0.85$T_g$[14]. Recent work by Qiu et al demonstrate how glass packing can influence chemical stability; denser glasses were found to be significantly more resistant towards reactions with atmospheric gases[55] and photodegradation[56]. For a reaction involving light induced conformational switching, it was found that a ≈ 1.3 % enhancement in glass density was accompanied by a factor of 50 improvement in photostability[57]. While these studies utilized model systems and reactions, they demonstrate how tight packing can slow reactions in the glassy state. In OLEDs, charge transport has been shown to promote chemical degradation[58,59]. Tight glassy packing prepared by PVD could slow these degradation processes, and consequently improve device lifetime. Interdiffusion of one layer into another in an OLED (e.g., interdiffusion of emissive layer into charge transport layer or vice-versa) can also degrade performance[60,61]; the dense and stable glasses (discussed in next section) prepared at ~0.85$T_g$ are more resistant towards interdiffusion[62].

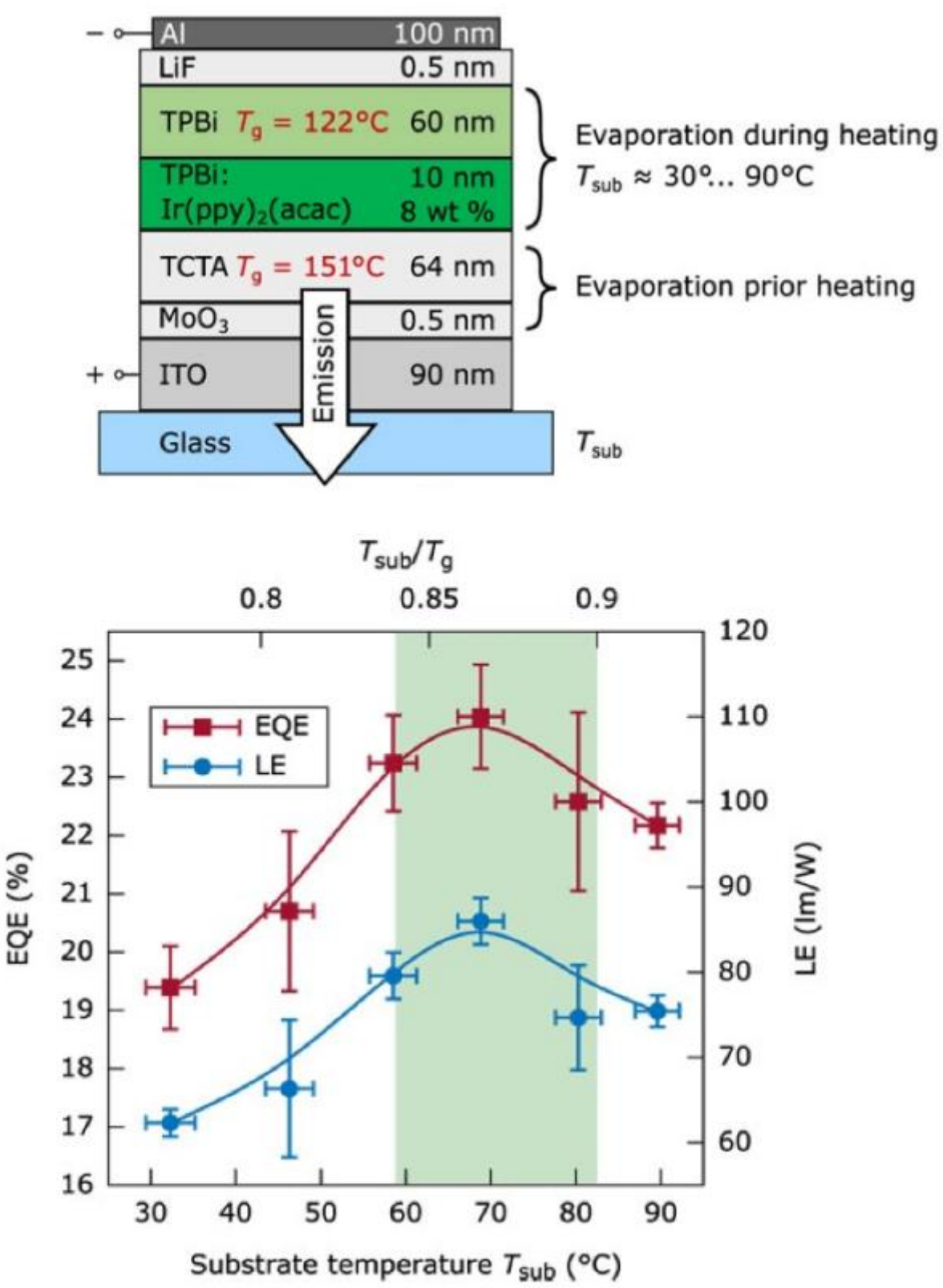

**Figure 6**: The device architecture of a green-light emitting OLED (organic light emitting diode), and its corresponding external quantum efficiency (EQE) and luminous efficacy (LE) as a function of the deposition temperature of the emitter and electron transport layers. The EQE and LE are evaluated at 100 cd /m$^2$. These measurements show that depositing at the optimal substrate temperature can significantly improve device performance. Measurements were performed by Reineke and co-workers. Reprinted from reference 14. © The Authors, some rights reserved;

**Stable glasses for organic electronics:** Given the evidence that it is beneficial to utilize stable glasses in OLED devices (**Figure 6**), in this section we discuss the expected generality of stable glass formation in vapor-deposited glasses of organic semiconductors.

Glasses of many organic molecules deposited at 80-85% of $T_g$, in addition to exhibiting higher density also exhibit enhanced thermal[63] and chemical stability[55], and suppressed motions in the glassy state[64,65]. Owing to their superior stability compared to glasses prepared from other routes, these materials are broadly referred to as "stable glasses". Stable glass formation is a common feature in vapor-deposited glasses of organic molecules, with more than 30 stable glass formers having been identified[32].

Shown in Figure 7 are the organic semiconductors that have been reported to form stable glasses. When these molecules are deposited at ~$0.85T_g$, they all form glasses with high kinetic stability[14,15,66]. In addition, vapor-deposited NPD and TPD have been shown to form high density glasses[15]. Conversely, we are not aware of any literature example of an organic semiconductor that does not form a stable glass when deposited near 0.85 $T_g$. We therefore expect that most organic semiconductors will form stable glasses when deposited at the appropriate deposition temperature. Molecules with a strong tendency to form intermolecular hydrogen-bonds may be an exception to this rule[67].

**Figure 7**: Molecular structures of organic semiconductors known to form stable glasses[14,15,66] when deposited at ~$0.85T_g$.

**New Horizons:** We identify the following as some of the outstanding challenges associated with vapor-deposited glasses of organic semiconductors:

*Controlling structure of PVD glasses at buried interfaces.* While well-established strategies exist to control the bulk structure of vapor-deposited glasses, little is known about the structure of vapor-deposited glasses at buried interfaces. Presently, even the length-scale over which the substrate can influence the structure of a vapor-deposited glass is poorly understood, with the first studies on this topic just recently appearing[68,69]. While dichroism studies by Yokoyama et al suggest the substrate can influence glass structure in films as thick as 100 nm[51], X-ray scattering measurements and molecular dynamics simulations suggest the substrate influences PVD glass structure for less than 10 nm[70,68].

Structure at buried interfaces plays an important role in organic electronic devices[71]. For OLEDs, structure at the interface of an inorganic electrode and an organic layer influences the charge injection barrier (into the organic layer). In OLEDs, charge transport across different organic layers is also likely to be influenced by structure at the organic-organic interface. Precisely controlling vapor-deposited glass structure at organic and inorganic buried interfaces may therefore have implications for OLED devices. Understanding structure at buried interfaces might also help us understand why, in PVD glasses of organic semiconductors, ≈10 nm films exhibit significantly different elastic modulus than thicker films (thickness > 20 nm)[72,73].

Resonant soft x-ray reflectivity, with its ability to depth profile molecular orientation, is a well suited technique to study vapor-deposited glass structure near interfaces[74]. NEXAFS spectroscopy on delaminated samples can also be applied to study structure at the interface of a vapor-deposited glass and inorganic electrode[75].

*Creating phase-separated morphologies with vapor deposition*. As shown in **Figure 3**, vapor-deposition is often used to prepare two-component glasses. While a homogeneous mixture is desired in an OLED device, for other applications including organic photovoltaics phase-separated morphologies are desirable[76]. The appropriate choice of molecules and deposition conditions could allow preparation of phase-separated morphologies via vapor-deposition; the enhanced mobility at the surface of the glass[77] would provide a kinetic route to phase separation when there is a thermodynamic driving force. Creating phase-separated materials and controlling domain sizes with deposition temperature and rate could expand the range of material properties accessible by vapor deposition for a given composition.

*Controlling energy transfer efficiency from host to guest by controlling vapor-deposited glass structure*. Efficient Förster resonance energy transfer (FRET) from the host to the guest molecules in the emitter layer is desirable for OLEDs[78]. It is well known that FRET is sensitive to the orientation of the donor molecule with respect to the acceptor[79,80]. **Figure 3** shows how the orientation of a dilute emitter in a host matrix is controlled by the deposition temperature relative to the glass transition temperature. Controlling the emitter orientation in a matrix with deposition temperature might provide an avenue to improve FRET rates in the emitter layer which in turn is likely to improve OLED performance. Control of FRET rates by manipulating glass structure has not been demonstrated yet, to the best of our knowledge.

*Simulation of charge mobility in vapor-deposited glasses*: Molecular dynamics and Monte-Carlo simulations can reliably predict the structure of vapor-deposited glasses[15,46,70]. The theoretical framework to simulate charge dynamics in molecular glasses has also been developed[81]. This opens opportunities for computationally studying charge carrier mobility in PVD glasses with different structure (but same composition). The ability to reliably simulate charge carrier mobility would enable screening a large range of molecules even before they are synthesized and deposited. A recent study by Riggleman and co-workers demonstrates how the properties of PVD glasses of a series of molecules, with systematic differences in chemical functionalization, can be investigated in-silico[82].

Initial publications showing the feasibility of calculating the charge carrier mobility in vapor-deposited glasses of organic semiconductors have already appeared and this area seems poised for rapid progress. Two recent studies computed the spatial autocorrelation of site energies for glasses of the organic semiconductor CBP(4,4′-Bis(N-carbazolyl)-1,1′-biphenyl)[70,83] vapor-deposited at different temperatures. Both studies found that CBP glasses in which molecules are more horizontally oriented exhibit longer spatial correlations of site energies which is favorable for charge transport. Future simulations could also potentially disentangle the relative roles of enhanced density and anisotropic glass structure on charge transport.

**Conclusion:** The structure and properties of vapor-deposited glasses of organic semiconductors can be controlled by the substrate temperature during deposition and deposition rate. Depositing under optimal conditions can result in significant enhancements in the external quantum efficiency and lifetime of an OLED device. In addition to allowing fabrication of better organic electronic devices, preparing glasses with different structures by PVD enables the development of structure-property relationships in the glassy state. Going forward, these structure-property relationships can be used to improve the broad variety of organic electronic devices that use organic glasses[84,85].

## ACKNOWLEDGMENT

We acknowledge financial support from the US Department of Energy, Office of Basic Energy Sciences, Division of Materials Sciences and Engineering, Award DE-SC0002161.We thank Camille Bishop, Kaichen Gu and Wolfgang Brütting for helpful conversations.